\documentstyle[titlepage,11pt]{article}
\newcommand{\capt}{\caption}

\newcommand{\gb}{Gay-Berne}

\newcommand{\beqn}{\protect\begin{equation}}
\newcommand{\eeqn}{\protect\end{equation}}
\newcommand{\fe}{flexoelectric effect}
\newcommand{\fc}{flexoelectric coefficient}
\begin{document}
\begin{center}
\bigskip
\bigskip
{\bf Molecular shape and flexoelectricity}\\
\bigskip
\bigskip
\bigskip
Jeffrey L. Billeter and
Robert A. Pelcovits$^\ast$\\
\bigskip
Department of Physics \\
Brown University \\
Providence, RI 02912\\
\end{center}
\vskip 3.0in
\begin{flushleft}
Running title: numerical simulations of flexoelectricity\\
\bigskip
\bigskip
$^\ast$ Corresponding author\\
Contact info: (401) 863-1432; (401) 863-2024 (FAX); pelcovits@physics.brown.edu
\end{flushleft}

\vfill
\eject

\begin{abstract}
We performed Monte Carlo simulations of systems of wedge--shaped objects formed from \gb\   ellipsoids joined to Lennard--Jones spheres. We studied two different wedge shapes, one more asymmetric than the other. The bend and splay flexoelectric coefficients were measured in the isotropic and smectic phases using linear response theory, and found to be negligibly small in the isotropic phase. We found a close connection between the properties of the intermolecular potential and the flexoelectric coefficients measured in the smectic phase. In particular, we found negligible bend coefficients for both shapes and a larger magnitude of the splay coefficient for the more prominent wedge, in accord with Meyer's original mechanism for flexoelectricity. The less prominent wedge produced a splay \fc\ with the opposite sign due to the attractive tail of the intermolecular potential and the relative narrowness of the molecular head. 
\end{abstract}
\break
\setcounter {page}{3}

\section{Introduction}

In the \fe\ \cite{meyer2} a 
director deformation produces
an electrical polarization, similar to the phenomenon of piezoelectricity
in solid crystals. The \fe\ was first proposed by Meyer \cite{meyer2} who considered asymmetric molecules either wedge-shaped with longitudinal dipole moments or ``banana'' shaped with transverse dipoles. 
In the absence of a director deformation, the packing of the molecules is similar to that of ellipsoidal shaped or rodlike molecules (the additional asymmetries of the
molecules have negligible effect) and the average polarization is zero. However, when a splay is imposed upon a
system of wedges or a bend upon a system of bananas, the preferred packing of the
molecules results in a net alignment of dipoles leading to an overall polarization
of the medium (see figures \ref{wedgesplayfig} and \ref{bansplayfig}). 
Alternatively, an applied electric field which aligns the dipoles induces a splay or bend in the appropriately shaped system---this is
sometimes called the inverse \fe. In either case, the net polarization ${\bf P}$
and the elastic deformations are related by the \fc s $e_{11}$ and $e_{33}$ introduced by Meyer through the following linear response relation: \begin{equation} \label{flexpoldefn}
{\bf P}=e_{11}\,{\bf \hat{n}}\left({\bf \nabla \cdot \hat{n}}\right)
+ e_{33}\,{\bf \hat{n}} \times \left({\bf \nabla \times \hat{n}}\right),\end{equation}
where $\bf\hat{n}$ is the director. The first term on the right hand side of this equation corresponds to splay flexoelectricity (relevant for wedges) and the second term to bend flexoelectricity (relevant for bananas).
\par
Subsequent to Meyer's work, Prost and Marcerou \cite{prost-mar} proposed a flexoelectric mechanism based on molecular quadrupoles requiring neither the shape
asymmetries nor the dipole moments of Meyer's original argument. Rather, molecular
quadrupoles allow uneven charge distributions leading to polarizations in any given
volume when a splay is imposed (see figure \ref{quadfig}). 
\par
Both the Meyer's dipole and Prost and Marcerou's quadrupole mechanisms have been observed 
experimentally \cite{skald-chuv,mar-prost1,mar-prost2,prost-persh,murt-rag-mad}.  Typically, the \fc s are measured
over a range of temperatures in the nematic phase and their variation compared with
that of the nematic order parameter $S$. Although it has been shown that for both the
dipole and quadrupole mechanisms there are contributions to the \fc s involving
several powers of $S$, Marcerou and Prost used the dominant contributions---$S$
for the quadrupole mechanism and $S^2$ for the dipole mechanism---in the analysis
of their experimental data \cite{mar-prost1}. They found flexoelectric coefficients proportional to $S$
for symmetric, non-polar
molecules, clearly demonstrating the quadrupole mechanism. Quadratic variation with
$S$ was seen for banana-shaped molecules with strong, transverse dipole moments.
\par
In the smectic $A$ phase, an additional \fc\ $e_{22}$ arises, representing the coupling
between the net polarization and variations in the smectic layer spacing \cite{prost-persh,degennes,four-dur}. Specifically, there is an additional contribution to the right--hand side of equation (\ref{flexpoldefn}) of the form: $e_{22}\,\frac{\partial^2u}
{\partial z^2}{\bf \hat{z}}$, where $u$ is the displacement of the smectic layers whose normals are parallel to the ${\bf \hat{z}}$ axis.
Prost and Pershan \cite{prost-persh} found this additional term 
to be negligible experimentally. The flexoelectricity of the smectic $C$ phase involves a total of 14 coefficients and is discussed in \cite{degennes}.
\par
A systematic method for calculating flexoelectric coefficients based on molecular shape and multipole properties would be very valuable. Mean--field theories \cite{prost-mar,straley,osipov,helf1,der-pet} do not consider the short-ranged fluctuations in molecular alignment which can be quite important. Computer simulations offer a way to assess the molecular origins of flexoelectricity. In particular simulations can focus strictly on Meyer's packing ideas without the complication of dipolar interactions which could lead to antiparallel alignment of side--by--side molecules. To date only one simulation study \cite{stelzer} of flexoelectricity has been carried out. In this study the flexoelectric coefficients were evaluated for wedge-shaped molecules interacting via a generalized Gay-Berne potential \cite{cleaver-care}. The molecules were modeled in this study (as well as in ours) by a Gay-Berne ellipsoid with a Lennard--Jones sphere attached near one end (see figure \ref{rodsphdeffig}). Two sets of parameters were considered in ref.\cite{stelzer}, one with a slightly more pronounced wedge shape than the other. The flexoelectric coefficients were evaluated using microscopic expressions based on density functional theory, and a larger value of $e_{11}$ was obtained for the more pronounced wedge. The bend flexoelectric coefficient was nearly zero, in agreement with Meyer's suggestion that bend flexoelectricity should not appear in a system of wedges.
\par
In this paper we consider a similar model but extend the study in several ways. First, we evaluate the flexoelectric coefficients using linear response theory and the fluctuation--dissipation theorem, which provide a computationally simpler and more direct means of evaluation compared with density functional theory.
Second, we consider two model sets of molecular parameters, one representing a molecule with significantly more asymmetry than either molecule considered in ref.\cite{stelzer}. By exploring the intermolecular potentials for our two sets of parameters we find that the less prominent wedges (similar to the less prominent ones considered in ref.\cite{stelzer}) prefer to align with their larger ends tilted {\it toward} each other, while the more prominent wedges prefer to align with their larger ends tilted {\it away} from each other. These opposing tendencies in turn lead to opposite signs of the splay flexoelectric coefficients, as one might expect from Meyer's model. As in ref. \cite{stelzer} we find a negligible value of the bend flexoelectric coefficient.
\par
The outline of this paper is as follows. In the next section we present the details of our  modeling of wedge--shaped molecules. Section III discusses the linear response theory used to measure the flexoelectric coefficients in our simulation. Our results are presented in section IV and we offer some concluding remarks in the final section.
\section{Molecular Shape Modeling}
\par
Using an approach similar to that of \cite{stelzer} we constructed a wedge--shaped molecule  from a standard \gb\
ellipsoid (or rod) with a sphere added near one end (see figure \ref{rodsphdeffig}). 
The net interaction potential between two wedge-shaped molecules labeled 1 and 2 then consists of four terms, namely
\beqn U_{tot}=U_{rod1-rod2}+U_{sphere1-sphere2}+U_{rod1-sphere2}+U_{sphere1-rod2}.\eeqn 
where $U_{rod1-rod2}$ is given by the original \gb\ potential \cite{gayberne}:
\begin{displaymath} U_{rod1-rod2}\left({\bf \hat{u}_1,\hat{u}_2,\bf {r}}\right)=4\varepsilon
\left({\bf \hat{u}_1,\hat{u}_2,\hat{r}}\right) \end{displaymath} \begin{equation} 
\times\left[\left\{\frac{\sigma_o}{r-\sigma\left({\bf \hat{u}_1,\hat{u}_2,\hat{r}}
\right)+\sigma_o}\right\}^{12}\!-\left\{\frac{\sigma_o}
{r-\sigma\left({\bf \hat{u}_1,\hat{u}_2,\hat{r}}\right)+\sigma_o}\right\}^6\right],
\label{GBpot}\end{equation} where ${\bf \hat{u}_1,\hat{u}_2}$
give the orientations of the long axes of rods $1$ and $2$, respectively, 
and ${\bf r}={\bf r_1}-{\bf r_2}$, with the centers of the rods located at positions ${\bf r_1}$ and ${\bf r_2}$. 
The parameter $\sigma\left({\bf \hat{u}_1,\hat{u}_2,\hat{r}}\right)$ is the 
separation between the rods at which the potential vanishes, and thus represents the
shape of the rods. Its explicit form is
\begin{eqnarray} \sigma\left({\bf \hat{u}_1,\hat{u}_2,\hat{r}}\right)&=&
\sigma_o\left[1-\frac{1}{2}\chi\left\{
\frac{\left({\bf \hat{r}\cdot\hat{u}_1}+{\bf \hat{r}\cdot\hat{u}_2}\right)^2}{
1+\chi\left({\bf \hat{u}_1\cdot\hat{u}_2}\right)}\right.\right. \nonumber\\ 
&&\left.\left.{}+
\frac{\left({\bf \hat{r}\cdot\hat{u}_1}-{\bf \hat{r}\cdot\hat{u}_2}\right)^2}{
1-\chi\left({\bf \hat{u}_1\cdot\hat{u}_2}\right)}\right\}\right]^{-1/2}, \label{sigma}\end{eqnarray}
where $\sigma_o=\sigma_\perp$ (defined below) and $\chi$ is
\begin{equation} \chi=\left\{\left(\sigma_\parallel/\sigma_\perp\right)^2-1\right\}/
\left\{\left(\sigma_\parallel/\sigma_\perp\right)^2+1\right\}. \end{equation} Here $\sigma_\parallel$
is the separation between two rods when they are oriented end-to-end with $U_{rod1-rod2}=0$, and
$\sigma_\perp$ is the corresponding separation when the two rods are side-by-side. 
The well depth $\varepsilon\left({\bf \hat{u}_1,\hat{u}_2,\hat{r}}\right)$,
representing the anisotropy of the attractive interactions, is
written as \begin{equation}\varepsilon\left({\bf \hat{u}_1,\hat{u}_2,\hat{r}}\right)
=\varepsilon_o\varepsilon^\nu\left({\bf \hat{u}_1,\hat{u}_2}\right)\varepsilon'^{\mu}
\left({\bf \hat{u}_1,\hat{u}_2,\hat{r}}\right),\label{varepsilonr}\end{equation} where
\begin{equation}\varepsilon\left({\bf \hat{u}_1,\hat{u}_2}\right)=\left\{1-
\chi^2\left({\bf \hat{u}_1\cdot\hat{u}_2}\right)^2\right\}^{-1/2},\label{eps}\end{equation} and
\begin{eqnarray} \varepsilon'\left({\bf \hat{u}_1,\hat{u}_2,\hat{r}}\right)&=&
1-\frac{1}{2}\chi'\left\{
\frac{\left({\bf \hat{r}\cdot\hat{u}_1}+{\bf \hat{r}\cdot\hat{u}_2}\right)^2}{
1+\chi'\left({\bf \hat{u}_1\cdot\hat{u}_2}\right)}\right. \nonumber\\ 
&&\left.{}+
\frac{\left({\bf \hat{r}\cdot\hat{u}_1}-{\bf \hat{r}\cdot\hat{u}_2}\right)^2}{
1-\chi'\left({\bf \hat{u}_1\cdot\hat{u}_2}\right)}\right\}, \label{epsprime}\end{eqnarray}
with $\chi'$ defined in terms of $\varepsilon_\parallel$ and $\varepsilon_\perp$, the end-to-end 
and side-by-side well depths, respectively, as
\begin{equation} \chi'=\left\{1-\left(\varepsilon_\parallel/\varepsilon_\perp\right)^{1/\mu}\right\}/
\left\{1+\left(\varepsilon_\parallel/\varepsilon_\perp\right)^{1/\mu}\right\}. \label{chiprime}\end{equation}
We measure all of our physical quantities in reduced units in terms of the energy scale  $\varepsilon_o$ and the length scale $\sigma_o$. 
For the adjustable parameters appearing in equations (\ref{varepsilonr}), (\ref{epsprime}) and (\ref{chiprime}), we used the values originally proposed by Gay and Berne \cite{gayberne}:  $\mu=2,\nu=1$, and
$\varepsilon_\perp/\varepsilon_\parallel=5$. Our choices for the rod shape parameters $\sigma_\parallel$ and $\sigma_\perp$ are discussed below.
\par
The interaction $U_{sphere1-sphere2}$ is given by the ordinary Lennard-Jones potential:
\begin{equation} U_{sphere1-sphere2}\left(\bf{r}\right)=4\varepsilon_o
\left\{\left(\frac{d}{r}\right)^{12}-\left(\frac{d}
{r}\right)^6\right\},
\end{equation}
where $d$ is the separation between the two spheres at which the potential $U_{sphere1-sphere2}$ vanishes; i.e., it is a measure of the diameter of the sphere. The relative position vector $\mathbf{r}$ is measured from the center of sphere 1 to the center of the sphere 2.

\par
The interaction between the rodlike part of one molecule and the sphere of the other molecule, $U_{rod1-sphere2}$, is given by 
a \gb\ potential generalized  to mimic the interaction between nonequivalent particles \cite{cleaver-care}.  The  range parameter $\sigma_{rs}\left(\mathbf{\hat{u}}_1,\mathbf{r}_{12}
\right)$ and the energy parameter $\varepsilon_{rs}\left(\mathbf{\hat{u}}_i,\mathbf{\hat{u}}_j\right)$
used in this potential are generalizations of the corresponding parameters in equations (\ref{sigma})
and (\ref{varepsilonr}), namely \cite{cleaver-care}
\beqn \sigma_{rs}\left({\bf \hat{u}_1,\hat{r}}\right)=\sigma_o^{rs}\left(1-\chi_{rs}\left(
\bf \hat{r}\cdot\hat{u}_1\right)\right)^{-1/2} \eeqn
and
\beqn \varepsilon_{rs}\left({\bf \hat{u}_1,\hat{r}}\right) = 1,\eeqn
with  $\chi_{rs}$ defined as 
\beqn \chi_{rs}=\frac{\sigma_\parallel^2-\sigma_\perp^{2}}
{\sigma_\parallel^2+d^2} ,\eeqn
and 
\beqn \sigma_o^{rs}=\frac{1}{\sqrt{2}}(\sigma_\perp^2 + d^2)^{1/2}.\eeqn
In the rod-sphere potential the relative position vector $\mathbf{r}$ is measured from the center of the rod to the center of the sphere.
\par
We performed simulations for two sets of molecular parameters:  a slightly wedge--shaped object with $\sigma_\perp=1.0,\;\sigma_\parallel=2.6,\;d=0.94$, and a more prominent wedge with parameters 
$\sigma_\perp=1.0,\;\sigma_\parallel=2.4,\;d=1.3$. In the former case the center of the sphere was located at a distance $D=1.3$ from the center of the rod, and at a distance $D=1.2$ in the latter case. In terms of the steric dipole moment $p^\ast= (4\pi/3)(d/2)^3 D$ introduced in \cite{stelzer}, our parameters correspond to values of $p^\ast$ of 0.565 and 1.38 respectively (in \cite{stelzer} systems with dipole moments of 0.524 and 0.662 were studied).
\par
To ascertain the effective shape of the composite rod--sphere molecule we computed equipotential contours (with contour values close to zero). The contour plots are shown for the two sets of parameters in figures \ref{swcontfig} and \ref{bwcontfig}, clearly indicating a wedge-like shape. To explore the interaction and local packing of the molecules we computed the depths of the potential wells of 
$U_{tot}$  for several different relative orientations of a pair of molecules with relative tilt $\theta$. The results for the two sets of molecular parameters are shown in figures \ref{swcurves} and \ref{bwcurves}.  For the purposes of comparison corresponding curves for the original \gb\ potential $U_{rod1-rod2}$ are shown in figure \ref{gbcurves}. Comparing figures \ref{swcurves} and \ref{bwcurves} with figure \ref{gbcurves}, we note that in the splay configuration the wedges do not exhibit the left--right tilt symmetry ($\theta \rightarrow \theta + 180^\circ$) seen in the \gb\ case. This result is not surprising: tilting the larger end of one wedge \textit{away} from the larger end of another wedge (corresponding to $\theta < 180^\circ$ in the figures) should yield a different energy than tilting the larger ends \textit{toward} each other. More importantly, for each set of wedge parameters there is an absolute minimum in the potential energy corresponding to a finite but small splay angle. For the more prominent wedge shape (figure \ref{bwcurves}) this angle is approximately $10^\circ$, i.e., the pair of wedges prefer to align with their larger ends tilted away from each other. For the less prominent wedge shape (figure \ref{swcurves}) the angle corresponding to the absolute minimum is nearly $360^\circ$, i.e., the wedges prefer to align with their larger ends tilted \textit{toward} each other. As we shall see below this difference yields flexoelectric coefficients of opposite signs for the two molecular shapes. While it might seem surprising at first glance that the less prominent wedges tilt toward each other, given the attractive interaction between two spheres and the fact that $d < \sigma_\perp$ in this case it is a reasonable result. For the more prominent wedges the larger repulsive core of the spheres leads to the expected tilting of the larger ends away from each other.
\par
While the absolute minima of the pair potentials for the two sets of wedge parameters correspond to nonzero splay, note that this state is only slightly preferred over the aligned state, $\theta=0^\circ$, for the parameters of figure \ref{swcurves} and the antiparallel aligned state, $\theta=180^\circ$, for the more prominent wedge of figure \ref{bwcurves}. There is also a substantial subsidiary potential minimum for antiparallel slightly wedge--shaped molecules. In our simulations of a system of 256 wedge--shaped molecules using both sets of molecular parameters we found no net spontaneous splay or electric polarization (assuming molecular dipole moments parallel to the long axis of the wedge). However, if an electric field were applied to the system of prominent wedges, we would expect the potential well corresponding to nonzero splay to become deeper relative to that of the antiparallel state $\theta=180^\circ$, leading to splay flexoelectricity. For the slightly wedge--shaped objects of figure \ref{swcurves}, the shallowness of the well corresponding to nonzero splay makes it less obvious that splay flexoelectricity will exist.  However, in our simulations we do in fact find splay flexoelectricity in this case, albeit smaller than in the case of the more prominent wedges and with the opposite sign.
\par
We now turn to a discussion of how we extract the flexoelectric coefficients from our simulations.

\section{Calculating flexoelectric coefficients}
To measure \fc s in our simulation, we used the linear response theory of 
Nemtsov and Osipov \cite{nem-os}. The flexoelectric coefficients in this method are related to the response function of the system to an orientational stress. Using the fluctuation-dissipation theorem the response function can be found from correlation functions of the polarization density and the orientational stress
tensor. The latter tensor 
is conjugate to the orientational strain which yields flexoelectricity. Thus, a calculation of the relevant correlation functions  yields the flexoelectric coefficients.
\par
Specifically, the splay and bend flexoelectric coefficients in the Nemstov--Osipov formalism are given by
\begin{eqnarray} \label{fceqns} e_{11}&=&-E_{\alpha \beta \gamma}e_{\mu \beta \gamma}
n_\alpha n_\mu /2 \nonumber\\
e_{33}&=&E_{\alpha \beta \gamma}e_{\alpha \beta \mu}n_\gamma n_\mu /2,\end{eqnarray}
where we use the summation convention over the Greek indices (summed over the coordinate directions $x,y$ and $z$). The tensor $e_{\alpha\beta\gamma}$ is the antisymmetric Levi--Civita tensor, while the antisymmetric tensor ${E_{\alpha\beta\gamma}}$ is the response function satisfying
\beqn P_{\alpha} = E_{\alpha\beta\gamma}\gamma_{ \beta\gamma}\eeqn
where $\gamma_{\alpha \beta}$ is the orientational strain tensor given by
\beqn \label{deformdefns}
\gamma_{\alpha \beta}=\frac{\partial \theta_\alpha}{\partial x_\beta}.\eeqn
Here $\theta_{\alpha}$ denotes the rotation angle of the director about the coordinate axis labeled by $\alpha$. For small director deformations  $\delta \mathbf{\hat {n}}$,  ${\bf \theta} \sim \sin{\theta} \sim \mathbf{\hat {n}} \times \delta \mathbf{\hat {n}}$, which then yields
\beqn  \gamma_{\beta \gamma}=
e_{\beta \mu \nu}n_\mu\frac{\partial n_\nu}{\partial x_\gamma}.\eeqn
Symmetry considerations show that ${E_{\alpha\beta\gamma}}$ has four independent components in the nematic and smectic A phases.
\par
Using the fluctuation--dissipation theorem the components of the response function $E_{\alpha\beta\gamma}$ are given by correlation functions of the orientational stress tensor and polarization:
\beqn E_{\alpha \beta \gamma}=-\frac{\beta}{V}
\left<\pi_{\beta \gamma}{\mathcal{P}}_{\alpha}\right>.\label{eabydefn}\eeqn
Here  $\beta = 1/k_B T$, $V$ is the volume of the system, ${\mathcal{P}}=
\sum_{i}\mbox{\boldmath$\mu$}_i$, ($\mbox{\boldmath$\mu$}_i$ is
the dipole moment of molecule $i$), and $\pi_{\alpha\beta}$ is the static orientational stress tensor given by
\beqn \pi_{\alpha \beta} = 
\frac{1}{2}\sum_{i\not=j}r_{ij\beta}\tau_{ij\alpha}.\label{pidefn}\eeqn
In the equation above ${\bf r}_{ij}$ is the relative position vector of molecules $i$ and $j$, and $\mbox{\boldmath$\tau$}_{ij}$ is the torque exerted by molecule $j$ on $i$.
\par
 In our simulations, then, we calculate the components $E_{\alpha \beta \gamma}$ of the response function by computing the correlation functions in equation (\ref{eabydefn})
as  time averages (or  MC cycle averages) and then calculate $e_{11}$ 
and $e_{33}$ from equation (\ref{fceqns}) (the
director is also computed during the simulations). We set the magnitude of the molecular dipole moment to unity; its direction is given by ${\bf \hat{u}}$, the long axis of the wedge. The torque on molecule $i$ due to our generalized potential $U_{tot}$ is given by \cite{luck-steph},
\beqn {\bf \tau}_i = \sum_{i\not=j}{\bf \tau}_{ij}= {\bf \hat{u}}_i \times \left(-\nabla_{\bf\hat{u}}U_{tot}\right).\eeqn

\par
\section{Results}
We performed constant temperature and pressure (with $P^\ast=P\sigma_o^3/\varepsilon_o = 10$) Monte Carlo simulations on systems of 256 wedges using the two sets of molecular parameters described in Section II. The systems were equilibrated for 250,000--500,000 cycles and cooled in dimensionless temperature steps of 0.1 (the dimensionless temperature is defined by $T^*\equiv k_BT/\varepsilon_o$). As in \cite{stelzer} we found a strong preference for the system to form a smectic A phase, even when we
removed the attractive part of the potential (in the ordinary GB system, removal of the attractive forces tends to stabilize the nematic phase \cite{mig-rull}). 
 The use of a different set of \gb\ parameters for the attractive portion of the potential \cite{berardi} (which enhance the stability of the nematic phase for a system of GB ellipsoids) did not stabilize the nematic phase in the present case of wedge--like molecules---another
indication that the repulsive core shape is the dominant factor in the possible formation of a nematic phase.
\par
Stelzer et al. \cite{stelzer} found that the nematic phase is absent for dipole moment $p^\ast=0.814$, though they were able to produce nematic phases over narrow temperature ranges for $p^\ast= 0.524$ and 0.662. Thus, our inability to produce a stable nematic phase for our molecules with $p^\ast=1.38$ is consistent with \cite{stelzer}, but our inability to produce a stable nematic for our less asymmetric shape with $p^\ast=0.565$ is not, and the reason for this discrepancy is not clear. We note that experimentally the splay and bend \fc s in the nematic and smectic phases  have
been found to be virtually identical \cite{prost-persh}, while the \fc  \ $e_{22}$ which appears in the smectic A phase but not the nematic is negligibly small. Thus, we proceeded to measure the splay and bend \fc s in the smectic phase using the method outlined in section III.   \par
 Interestingly, the
smectic layers showed a distinct domain structure (see figure \ref{domainsfig})
characterized by alternating regions of parallel molecular alignment (compare
with the \gb\ smectic also shown in figure \ref{domainsfig}).
This structure seems consistent with the idea that the parallel wedges prefer splay;
splay occurs in each domain while alternation of the alignment between domains
maintains the flat, parallel smectic layers. 
\par
Data for the small ($p^\ast=0.565$) and large ($p^\ast=1.38$) wedges and for \gb\ ellipsoids are shown in Table \ref{table1}. The \gb\ ellipsoids interact via the rod potential equation (\ref{GBpot}) with parameters $\sigma_\perp =1.0, \sigma_\parallel=3.0, \mu=2$, and $ \nu=1$. We note that the values of the flexoelectric coefficients for the system of ellipsoids are zero to within the computed error; thus in accord with Meyer's ideas this system of symmetric objects does not exhibit flexoelectricity. In the isotropic phase both the large and small wedges also exhibit no flexoelectricity. However, in the smectic phases of both wedges splay flexoelectricity appears.
Note the difference in sign between the
$e_{11}$ data in the smectic phase for the large and small wedges, consistent with their opposite
splays. The actual values for the signs also are as expected: considering
again figure \ref{wedgesplayfig} with the director $\mathbf{\hat n}$ taken to point upwards,
the splay shown is then positive and the resulting polarization is parallel
to $\mathbf{\hat n}$, implying that $e_{11}>0$ (recall equation (\ref{flexpoldefn}));
this is the case for the large wedge. For the small wedge, the splay is negative
but the polarization is still parallel to $\mathbf{\hat n}$, implying $e_{11}<0$
as observed. The magnitude of $e_{11}$ is also larger for the large wedge consistent with Meyer's excluded volume mechanism for flexoelectricity. The
average values for $e_{33}$, the bend flexoelectric coefficient, are much smaller in the smectic phases than the corresponding $e_{11}$ values and are zero to within the computed error. We also note that the bend coefficients varied in sign over the course of the simulation whereas the values of $e_{11}$ maintained a consistent sign during the runs.

\section{Conclusion}
By simulating a system of wedge--like molecules formed from Gay--Berne ellipsoids and Lennard--Jones spheres we have explored some of the molecular origins of flexoelectricity. We measured both the bend and splay flexoelectric coefficients using linear response theory, which yields the coefficients in terms of correlation functions of the molecular torque and orientation vector. We studied wedges with two different parameterizations, and found a close connection between the properties of the intermolecular potential and the flexoelectric response of the system. In particular, wedge-shaped molecules do not produce a measurable bend \fc, and a more prominent wedge-shaped object produces a larger splay \fc, in accord with Meyer's original ideas on the origins of flexoelectricity. In the case of the less prominently shaped wedge we have obtained a splay \fc\ with the opposite sign to that of the more prominent wedge due to the attractive tail of the intermolecular potential and the relative narrowness of the molecular head.
\section*{Acknowledgements}
We are grateful to Prof. G. Crawford for helpful discussions. Computational work in support of this research was performed at the
Theoretical Physics Computing Facility at Brown University. Our work was 
supported by the National Science Foundation under grants 
DMR9528092 and DMR9873849.

\vfill\eject
\begin{figure}
\caption{(a) Wedges with longitudinal dipoles under normal nematic conditions.
There is no splay and no net polarization. (b) Under an applied splay,
the preferred wedge alignment results in a net polarization. Alternatively,
an applied field induces a splay due to the wedge shape of the molecules.}
\label{wedgesplayfig}\end{figure}
\begin{figure}
\capt{(a) ``Bananas'' with transverse dipoles under normal nematic conditions.
There is no bend and no net polarization. (b) Under an applied bend,
the preferred banana alignment results in a net polarization. Alternatively,
an applied field induces a bend due to the banana shape of the molecules.}
\label{bansplayfig}\end{figure}
\begin{figure}
\capt{Quadrupoles with a splay imposed. Within the central ``layer'',
plus charges from above are allowed to enter, while plus charges below
are expelled, leading to a net polarization upwards.}
\label{quadfig}\end{figure}
\begin{figure}
\capt{Illustration of the basic geometric parameters of the wedge--shaped molecule composed of a rod and sphere.}
\label{rodsphdeffig}\end{figure}
\begin{figure}
\capt{Potential energy contours showing the shape of the rod-sphere composite with parameters $\sigma_\perp=1.0,\;\sigma_\parallel=2.6,\;d=0.94$: (a) side
view showing wedge-like asymmetry and (b) top view showing axial symmetry. 
The top view contours are not completely circular due to a finite number of sample points.
The side view data was generated by fixing one wedge in the $yz$ plane as shown,
while a second wedge pointing up out of the plane and with its narrow end just
touching the plane ``scanned'' across the first wedge. For the top view data,
both wedges were parallel; one remained fixed while the other was moved around 
the first in the same plane.}
\label{swcontfig}\end{figure}
\begin{figure}
\capt{Potential energy contours showing the shape of the rod-sphere composite for a more
prominent wedge with parameters $\sigma_\perp=1.0,\;\sigma_\parallel=2.4,\;d=1.3$.}
\label{bwcontfig}\end{figure}
\begin{figure}
\capt{Minimum well depths (calculated as a function of molecular separation) for
molecular parameters $\sigma_\perp=1.0,\;\sigma_\parallel=2.6,\;d=0.94$ for
various relative orientations as shown: (a) splay, (b) twist and
(c) bend configurations.  }
\label{swcurves}\end{figure}
\begin{figure}
\capt{Same as figure \ref{swcurves} but for the more prominent wedges shown
above (figure \ref{bwcontfig}) with parameters $\sigma_\perp=1.0,\;\sigma_\parallel=2.4,\;d=1.3$.}
\label{bwcurves}\end{figure}
\begin{figure}
\capt{Same as figure \ref{swcurves} but for \gb\ molecules with parameters $\sigma_\perp=1.0,\;\sigma_\parallel=3.0.$}
\label{gbcurves}\end{figure}
\begin{figure}
\capt{View of the molecular configurations in the smectic layers for
(a) wedges and (b) standard Gay-Berne ellipsoids. The arrows indicate
the direction of the orientation vectors $\mathbf{\hat u}$ whose heads
correspond to the wide ends of the wedges. The domains in the wedge case
can be clearly seen.}
\label{domainsfig}
\end{figure}

\begin{table}
\caption{Flexoelectric coefficients for the small ($p^\ast=0.565$) and large ($p^\ast=1.38$) wedges and for \gb\ ellipsoids. Data for the small wedge was obtained in the isotropic phase at $T^\ast=3.5$ and in the smectic phase at $T^\ast=2.9$; corresponding values for the large wedge were 2.5 and 1.9. Data for the \gb\ ellipsoids was obtained in the smectic phase at $T^\ast=0.745$ and $P^\ast=2.5$. The nematic order parameter in each of the smectic phases was 0.9.} 
\begin{center} 
\begin{tabular}{||c|c|c||}
\hline
  & $e_{11}$ & $e_{33}$ \\ 
\hline 
 small wedge, isotropic phase & $0.367 \pm 0.522$& $-0.394 \pm 0.383$\\ 
large wedge, isotropic phase & $-1.39 \pm 1.15 $ & $ 0.148 \pm 0.251$\\
Gay-Berne ellipsoids, nematic phase & $-0.394 \pm 0.581 $ & $ -0.0835 \pm 0.117$\\
\hline
small wedge, smectic phase & $-2.08 \pm 0.211$ & $-0.061 \pm 0.054$\\
large wedge, smectic phase & $13.6 \pm 0.052 $ & $ -0.005 \pm 0.012$\\
\hline \end{tabular} \end{center} \label{table1}
\end{table}
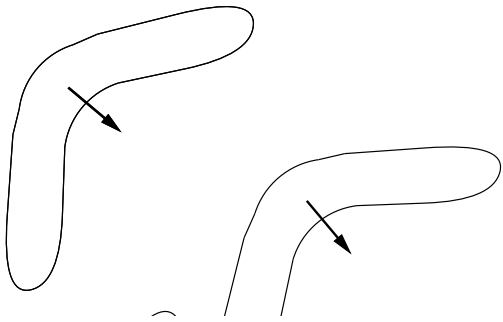
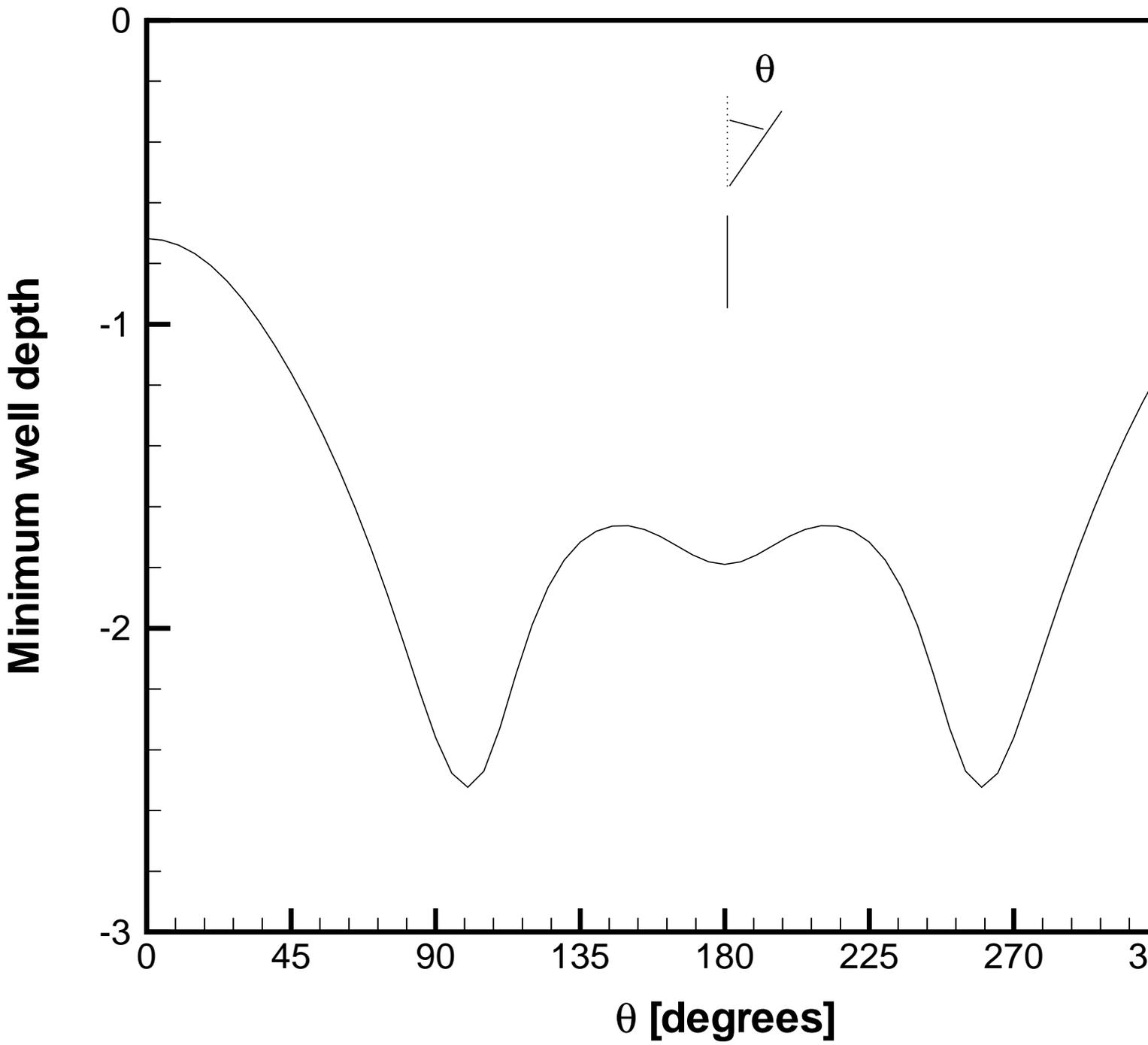
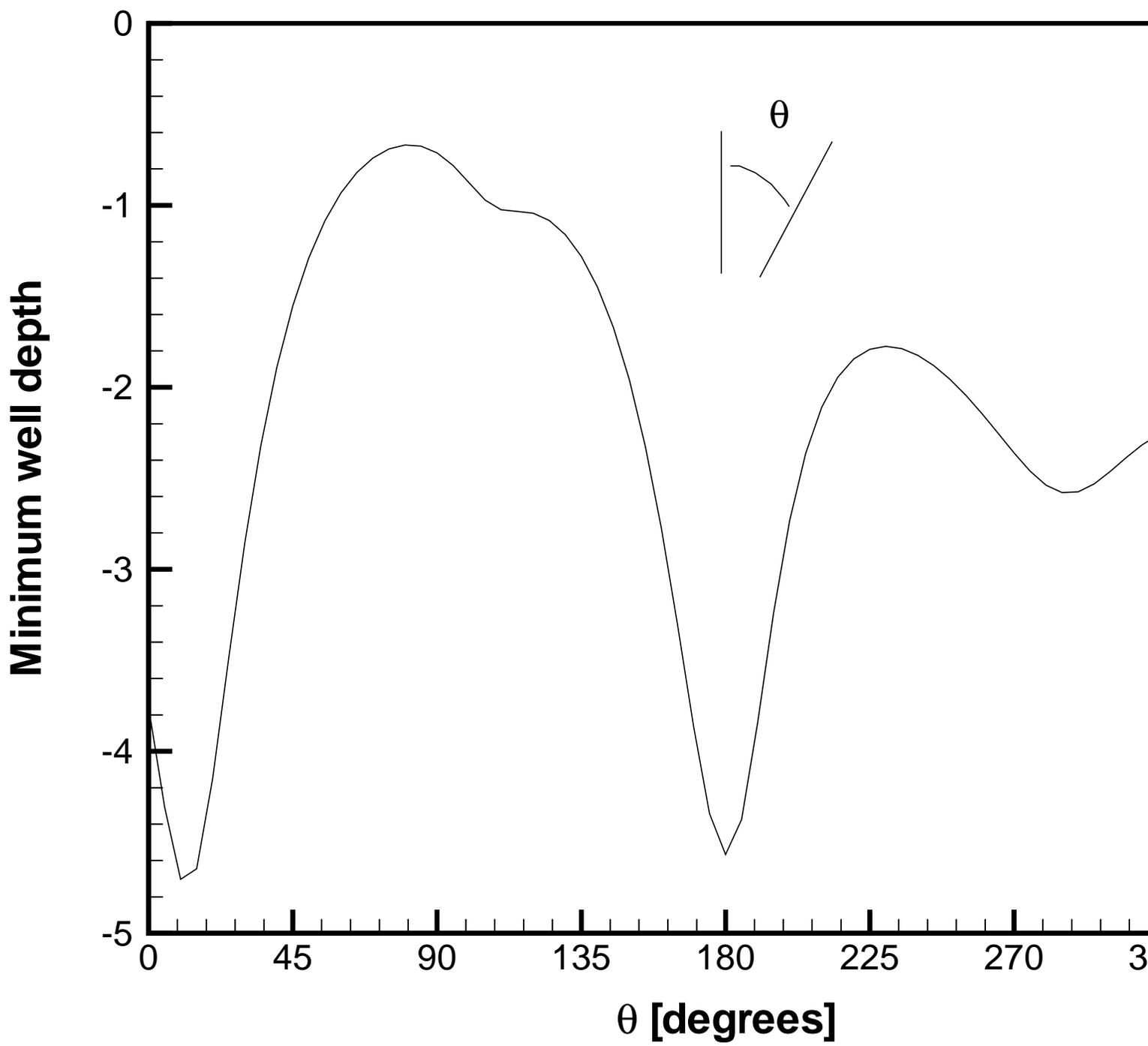
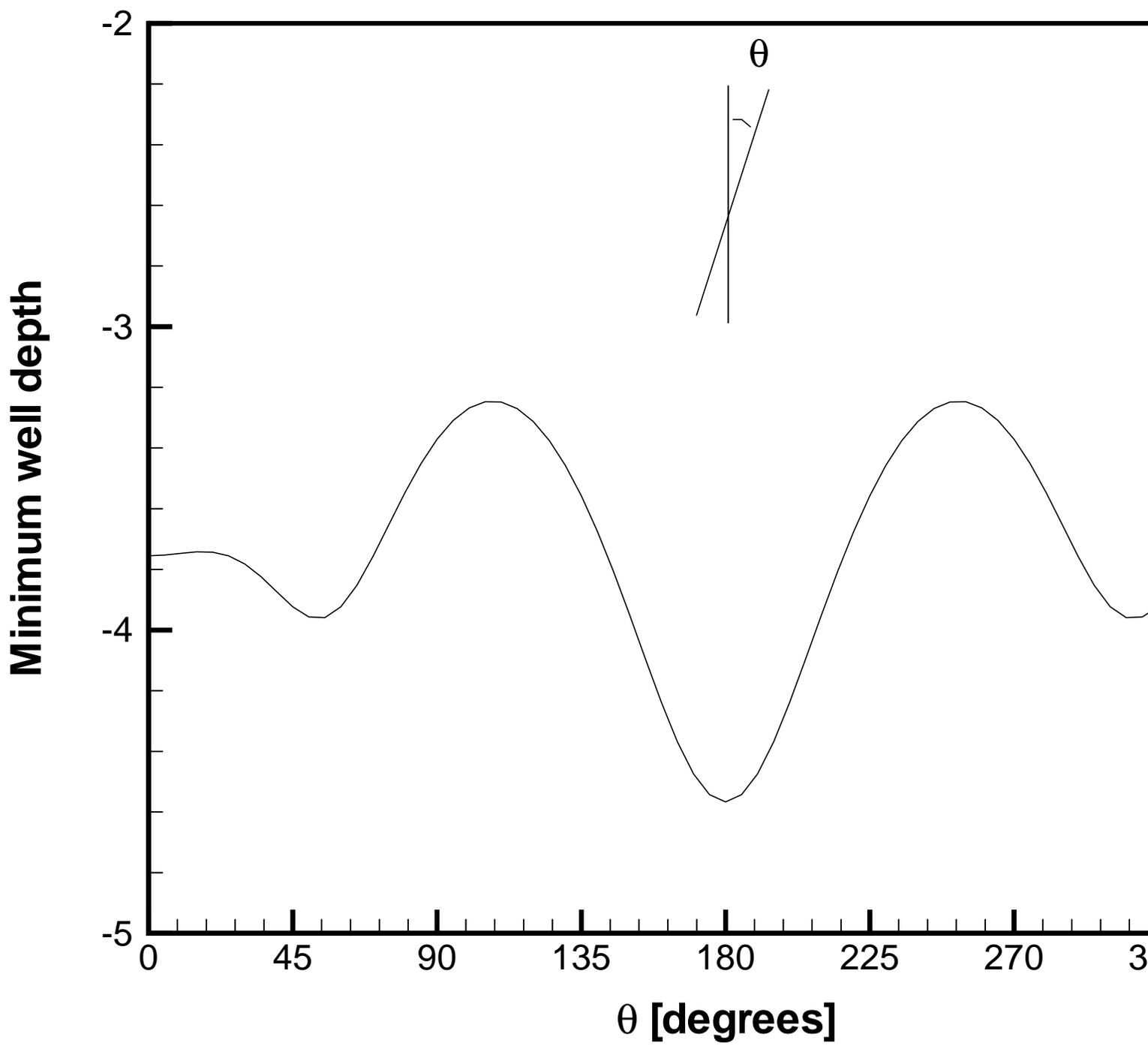
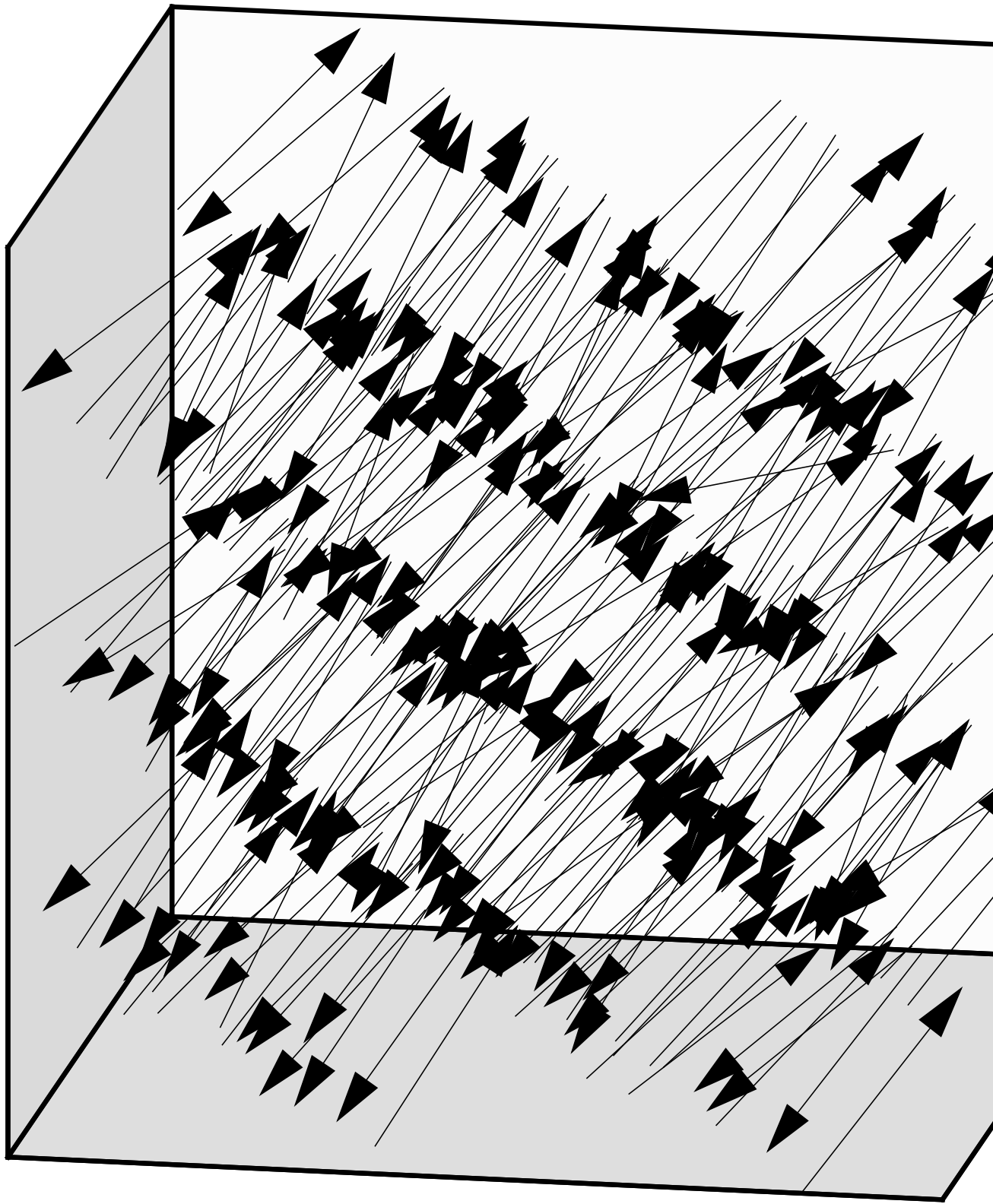
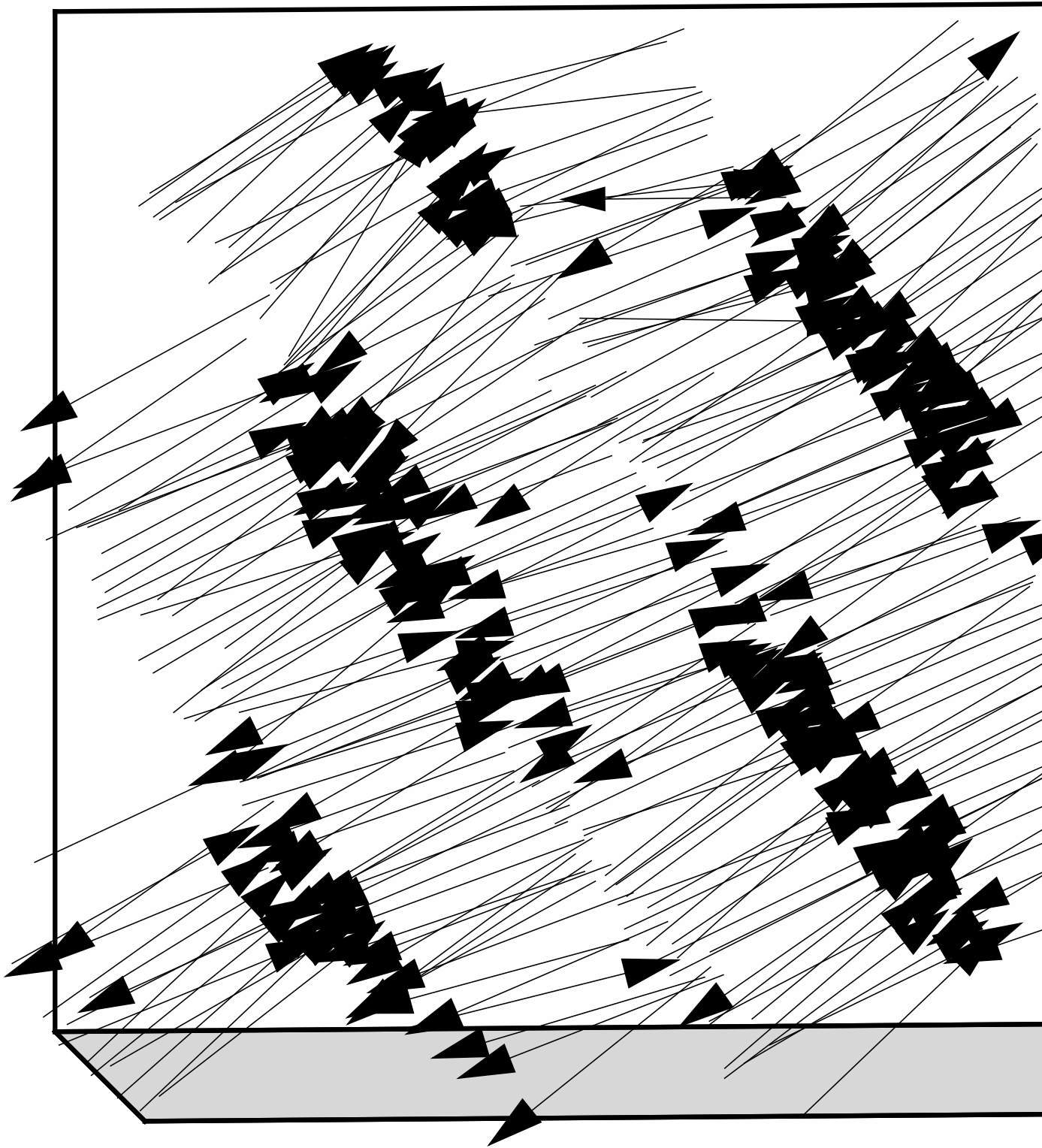
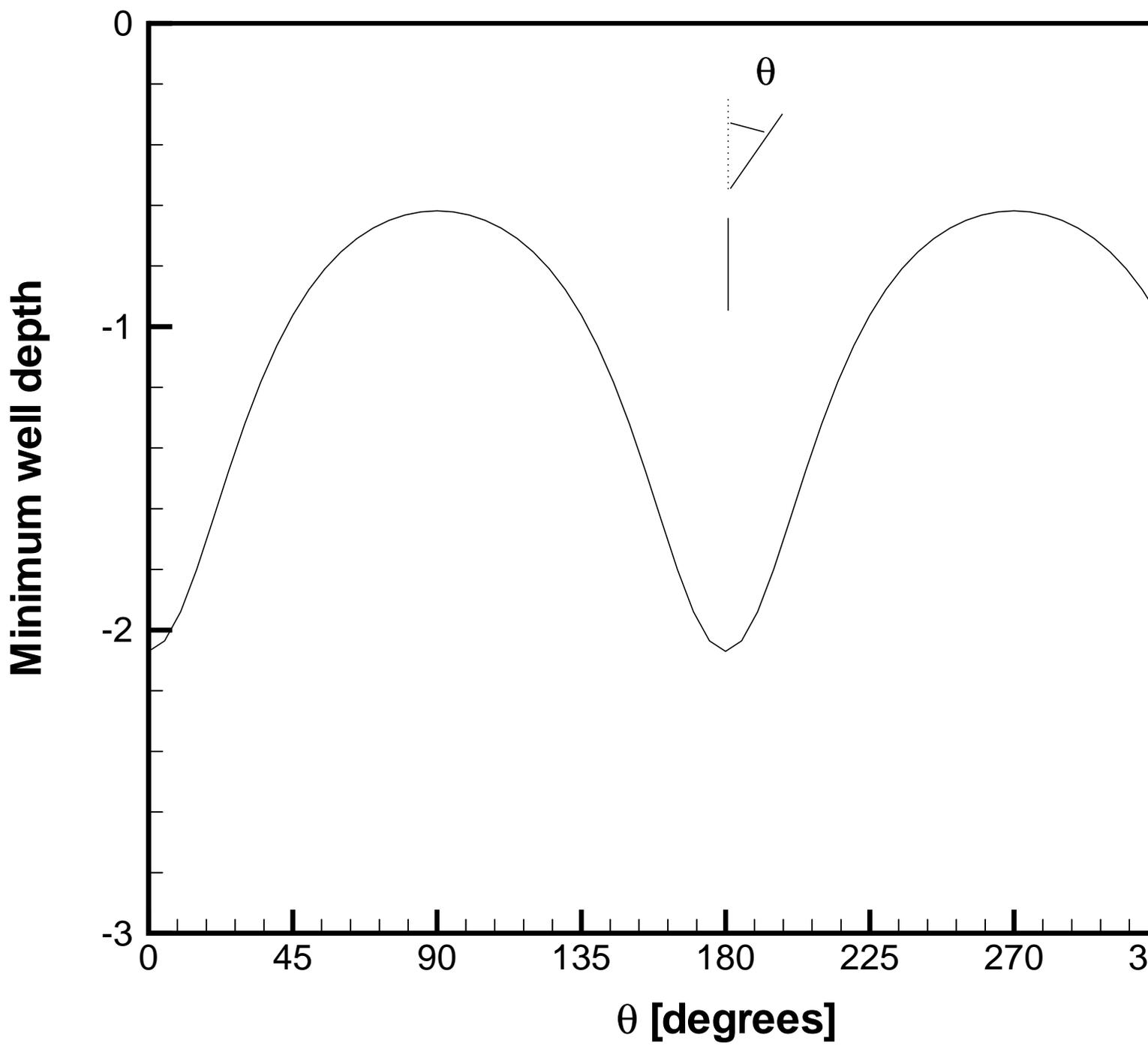
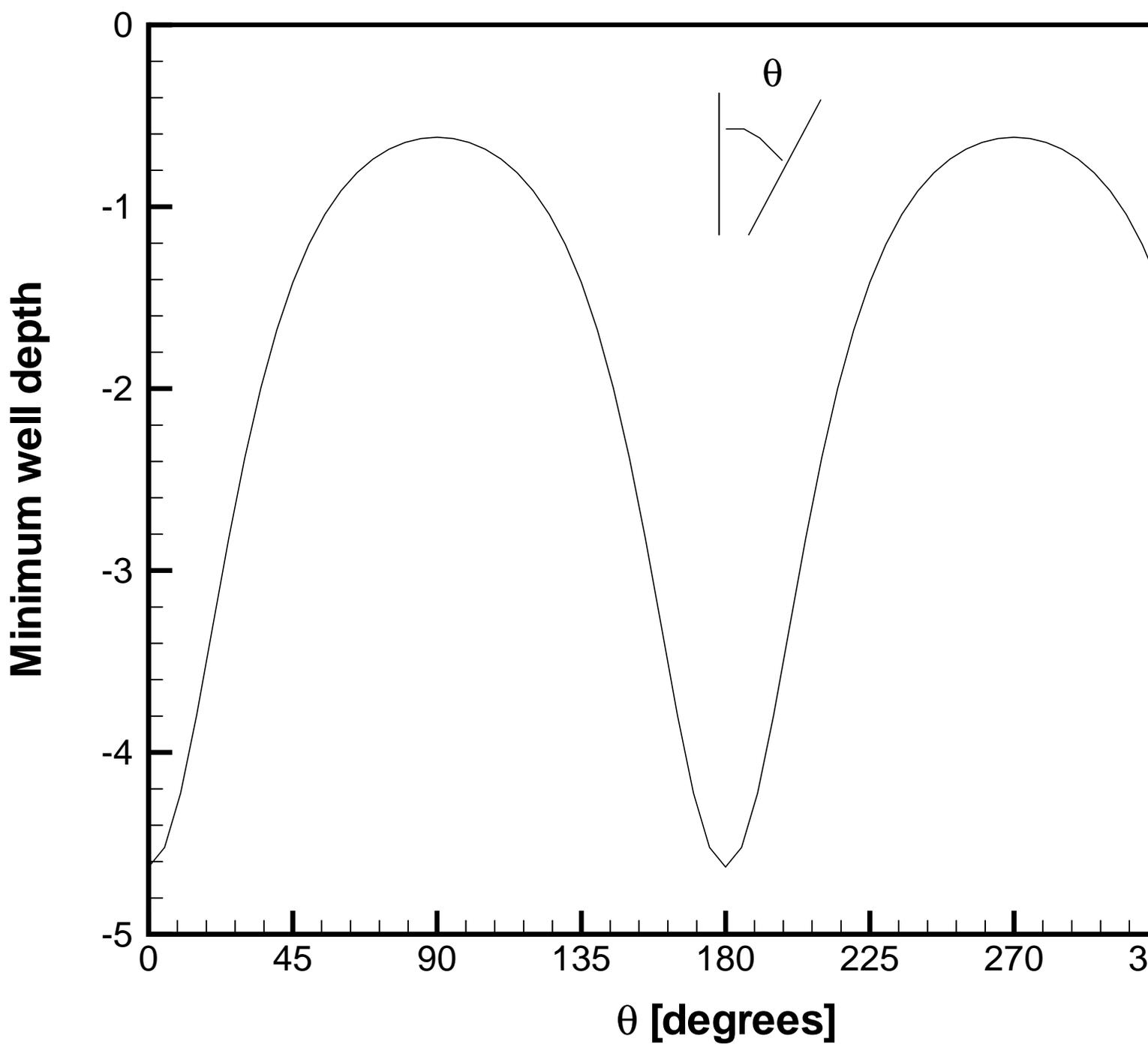
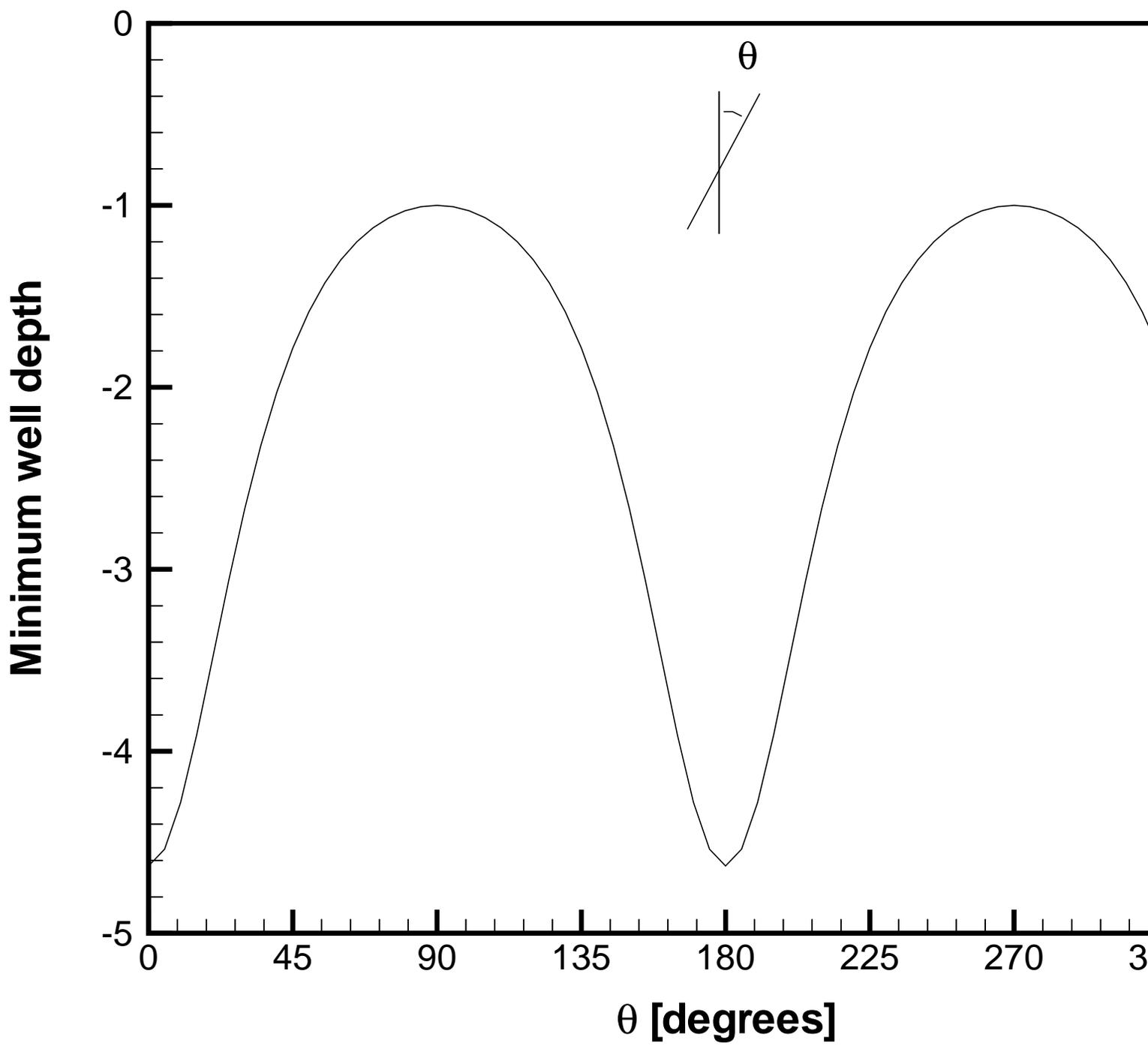
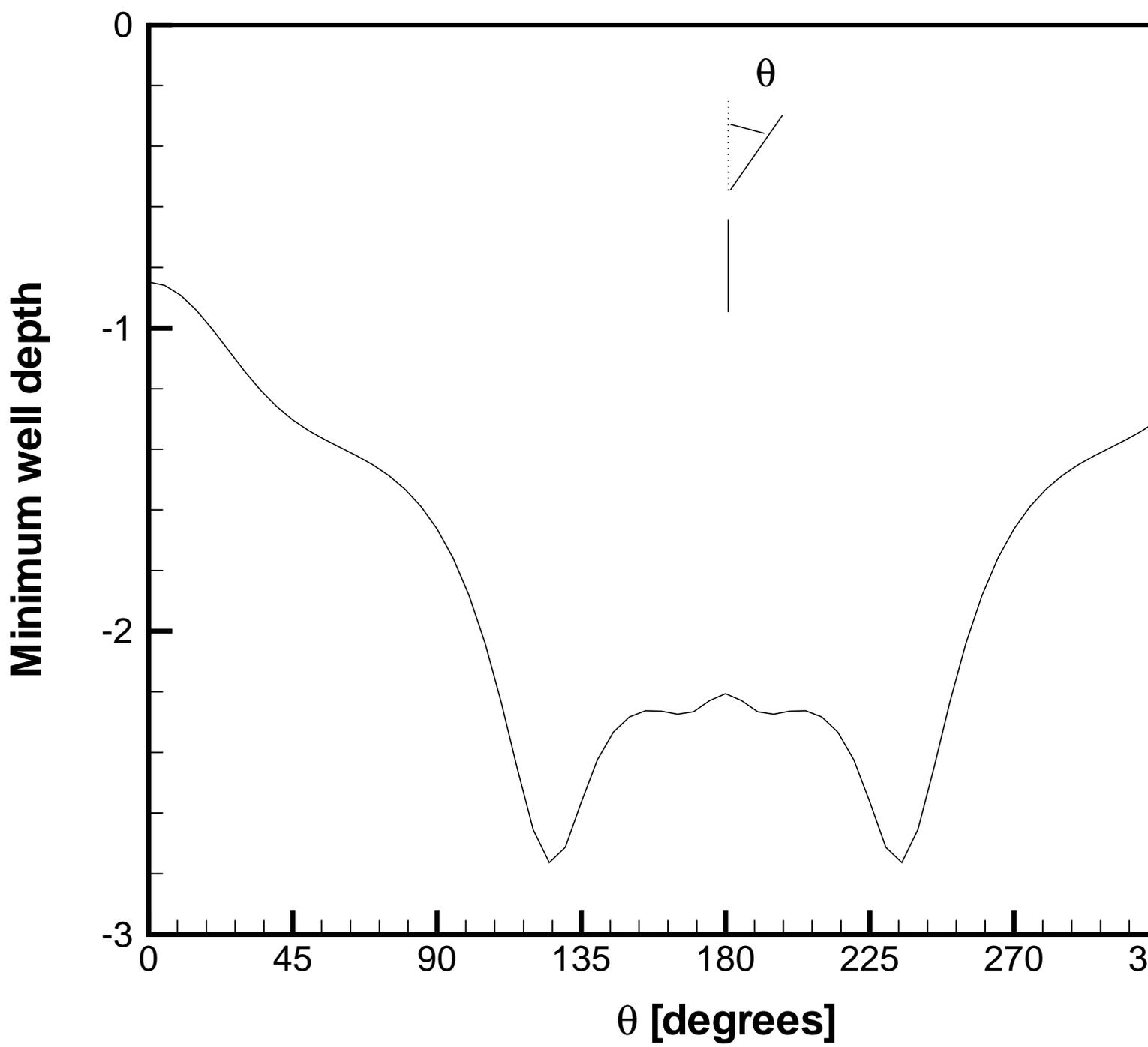
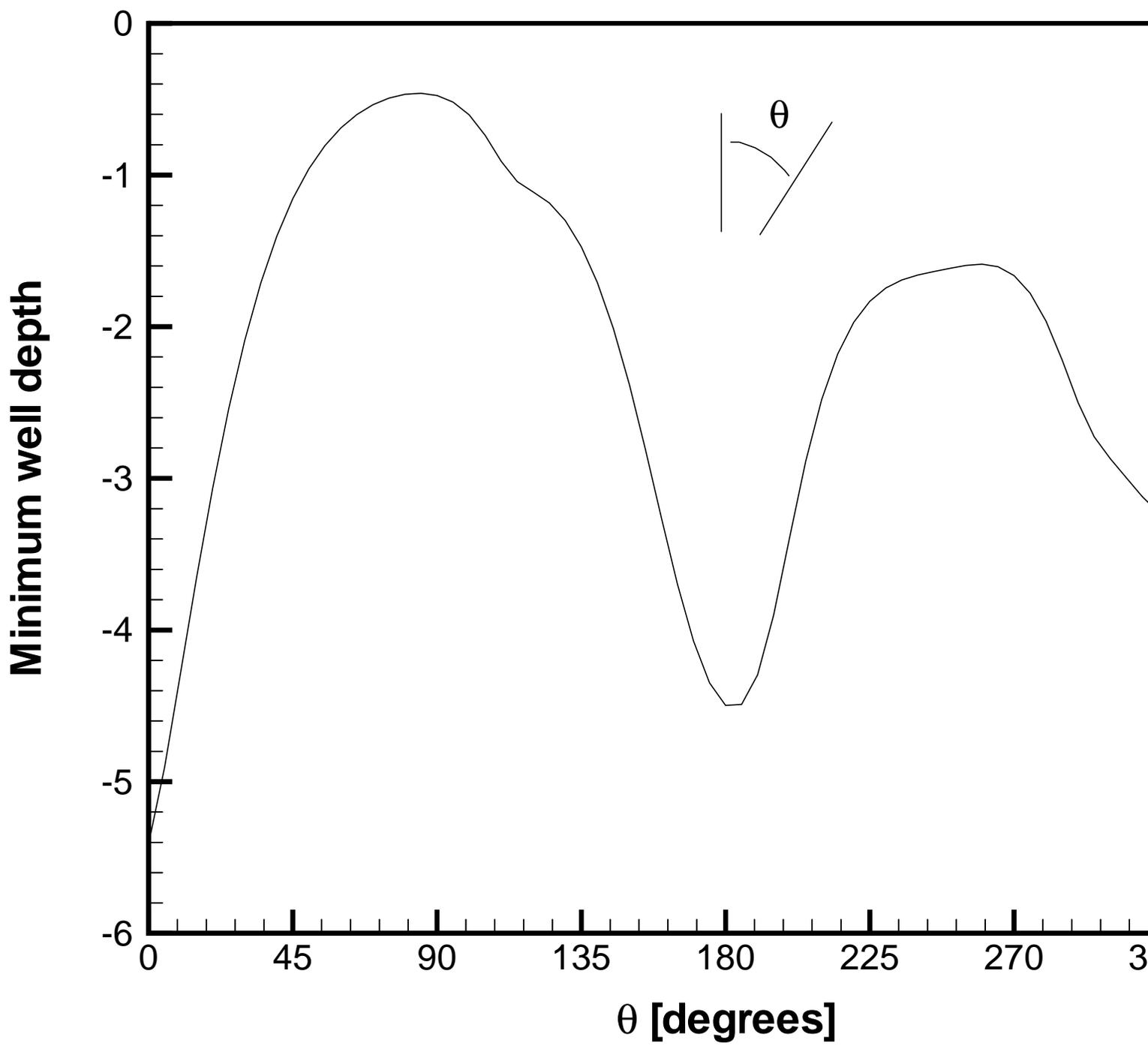
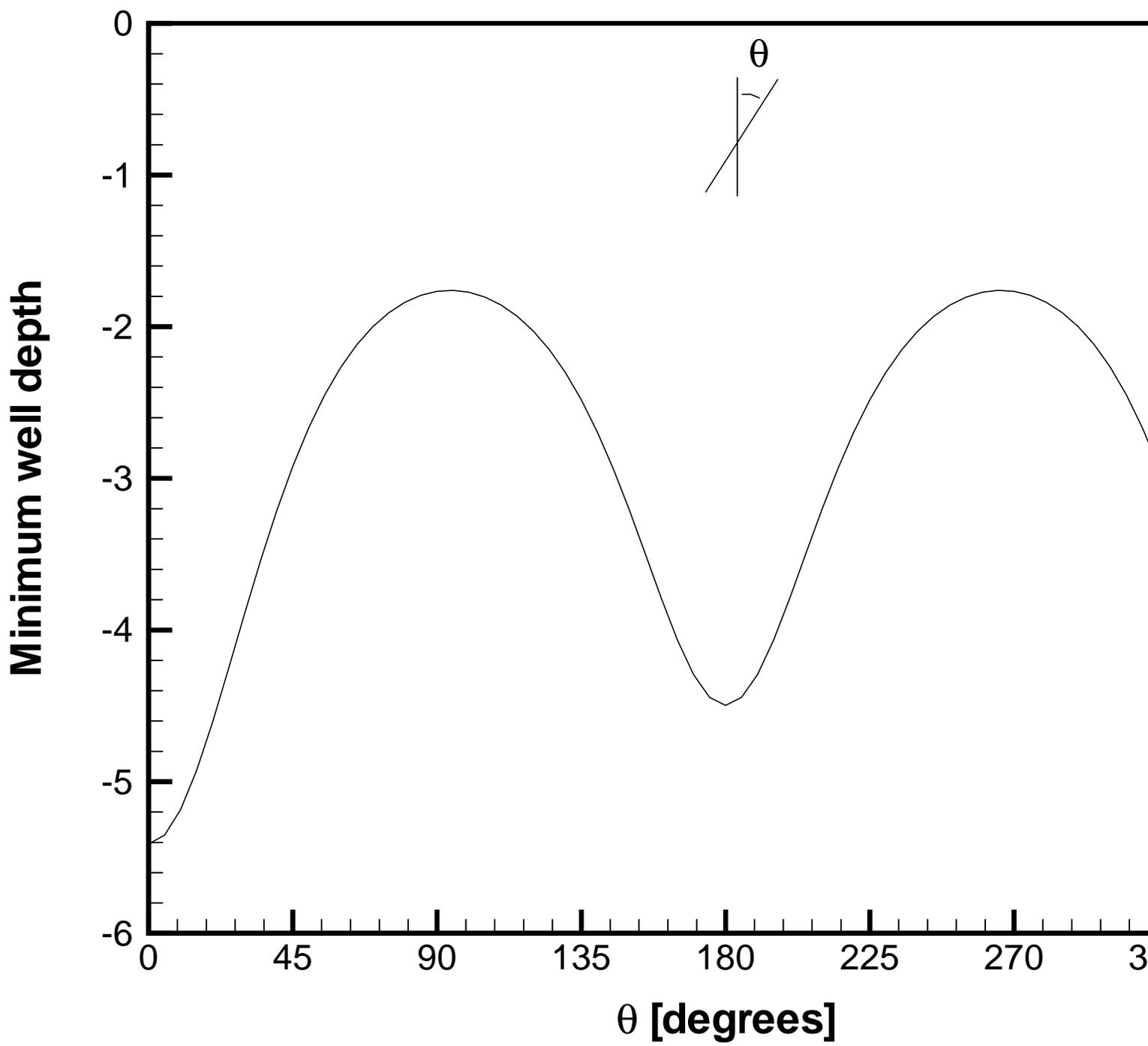
\end{document}